\def \SAIT #1 #2 {{\em Mem.\ Soc.\ Astron.\ It.\/} {\bf #1}, #2}
\def \MESS #1 #2 {{\em The Messenger\/} {\bf #1}, #2}
\def \ASTRNACH #1 #2 {{\em Astron. Nach.\/} {\bf #1}, #2}
\def \AAP #1 #2 {{\em Astron. Astrophys.\/} {\bf #1}, #2}
\def \AAL #1 #2 {{\em Astron. Astrophys. Lett.\/} {\bf #1}, L#2}
\def \AAR #1 #2 {{\em Astron. Astrophys. Rev.\/} {\bf #1}, #2}
\def \AAS #1 #2 {{\em Astron. Astrophys. Suppl. Ser.\/} {\bf #1}, #2}
\def \AJ #1 #2 {{\em Astron. J.\/} {\bf #1}, #2}
\def \ANNREV #1 #2 {{\em Ann. Rev. Astron. Astrophys.\/} {\bf #1}, #2}
\def \APJ #1 #2 {{\em Astrophys. J.\/} {\bf #1}, #2}
\def \APJL #1 #2 {{\em Astrophys. J. Lett.\/} {\bf #1}, L#2}
\def \APJS #1 #2 {{\em Astrophys. J. Suppl.\/} {\bf #1}, #2}
\def \APSS #1 #2 {{\em Astrophys. Space Sci.\/} {\bf #1}, #2}
\def \ASR #1 #2 {{\em Adv. Space Res.\/} {\bf #1}, #2}
\def \BAIC #1 #2 {{\em Bull. Astron. Inst. Czechosl.\/} {\bf #1}, #2}
\def \JSQRT #1 #2 {{\em J. Quant. Spectrosc. Radiat. Transfer\/} {\bf #1}, #2}
\def \MN #1 #2 {{\em Mon. Not. R. Astr. Soc.\/} {\bf #1}, #2}
\def \MEM #1 #2 {{\em Mem. R. Astr. Soc.\/} {\bf #1}, #2}
\def \PLR #1 #2 {{\em Phys. Lett. Rev.\/} {\bf #1}, #2}
\def \PASJ #1 #2 {{\em Publ. Astron. Soc. Japan\/} {\bf #1}, #2}
\def \PASP #1 #2 {{\em Publ. Astr. Soc. Pacific\/} {\bf #1}, #2}
\def \NAT #1 #2 {{\em Nature\/} {\bf #1}, #2}

\documentstyle{memsait}
\input epsf.sty
\begin{opening}
\date{} 

\title{Do zero metal intermediate mass stars experience thermal pulses?}
  
\author {Inma Dom\'{\i}nguez$^1$, Oscar Straniero$^2$, Marco Limongi$^3$, Alessandro Chieffi$^4$}
\institute{$^1$Dpto. de F\'{\i}sica Te\'orica y del Cosmos, Universidad de Granada,
18071 Granada, Spain\\ $^2$Osservatorio Astronomico di Collurania, 64100 Teramo, Italy\\ $^3$Osservatorio di Roma,
Via Osservatorio 2, I-00040 Monte Porzio (RM), Italy\\$^4$Istituto di Astrofisica Spaziale (CNR), Roma, Italy}

\date{} 
\end{opening}

\begin{document} 
\oddpagefooter{}{}{} 
\evenpagefooter{}{}{} 
\
\bigskip

\begin{abstract}
We have studied the evolution of intermediate mass (M$\ge$ 5M$_\odot$) 
zero metal (Z=0) stars with particular attention to the AGB phase. 
At variance with previous claims we find that these stars
experience thermal instability (the so called thermal pulses).
The critical quantity which controls the onset of a thermally pulsing phase 
is the amount of CNO in the envelope during the AGB. 
For these stars the central He burning starts in the blue side 
of the HR diagram and the 1$^{st}$ dredge up does not take place. Then the envelope
maintains its initial composition up to the beginning of the AGB phase.     
However, during the early AGB the $2^{nd}$ dredge-up occurs and fresh He and CNO elements
are engulfed in the  convective  envelope. We find that in    
stars with M$\ge$ 6M$_\odot$ the resulting 
amount of $^{12}C$ is large enough 
to sustain a {\it normal} CNO burning within the H shell and consequently the 
star enters the usual thermal pulse phase.    
In the 5M$_\odot$ model, owing to the lower $^{12}C$ enhancement in the envelope
after the  $2^{nd}$ dredge-up, the He burning shell
suffers weak thermal instabilities. 
9 of these thermal oscillation are needed before the He burning luminosity
reaches $3\cdot10^5 L_{\odot}$ and a first convective shell
develops in between the two burning shells. Later on a second convective shell
forms at the base of the H rich envelope. This convective zone cross the H/He discontinuity and
partially overlaps the previous one, 
dredging up fresh $^{12}C$. 
After a huge H flash, a quiescent CNO burning settle on. From this moment       
a thermal pulse phase starts, which is
very similar to the one experienced by the more massive models. 

\end{abstract}

\section {Introduction}

Following the standard homogeneous Big Bang nucleosynthesis scenario, the primordial Universe
were mainly composed by  H, $^4$He and few other light elements (lighter than Carbon). 
Heavier elements, from Carbon to Uranium, were built inside stars.  

Thus, the first generation of stars, those having a zero metal composition,
constitutes the 
initial step of the chemical evolution of the Universe after the big bang. It is
usually referred as Population III ( or Pop III). 
Several subjects of astrophysical studies require a good knowledge
of the properties of this Pop III: 
collapsing clouds, structure and galaxy formation, re-ionization of the universe,
chemical and dynamical evolution, and so on.
However, due to the high uncertainties and lack of direct 
observations, few studies have been devoted to the comprehension of the
evolutionary properties of these ancestors of the present stellar populations.

\begin{figure}
\epsfysize=6cm 
\epsfxsize=8cm 
\hspace{2cm}\epsfbox{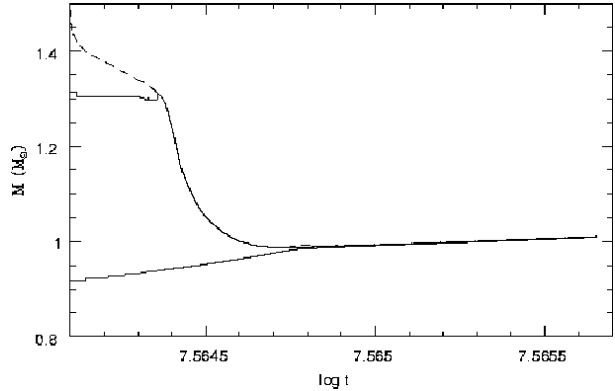} 
\caption[h]{Variation  of the positions of the H and He burning shells during the AGB for the 
7 M$_\odot$ (solid lines). The position of the bottom of the convective envelope is also
shown (dashed line). Note the 
significant penetration of the convective envelope into the He core during the E-AGB.}
\end{figure}

Great observational efforts have been done to identify 
very metal poor stars and by now around 70 of these objects have been selected 
(Bond 1981, Bessel and Norris 1984, Beers, Preston and Shectman 1992; Ryan, Norris and Bessel 1991; 
Ryan, Norris and Beers 1996;  Sneden et al. 1994, Primas et al. 1994, 
Ryan et al. 1991, Carney and Peterson 1991, Molaro and Castelli 
1990, Molaro and Bonifacio 1990). 

In principle, stars could be formed from  
an initial massive object deprived of metals. In such a case the molecular H 
provides an efficient cooling of the gas (Palla, Salpeter and Stahler, 1983), so that the  
Jean mass drops well below the stellar values. Several studies obtained a pre-galactic
population composed by  
rather massive objects, which could undergo a subsequent fragmentation leading
to the formation of stars or black holes (see  Tegmark et al. 1997 and references therein).      
If stars are formed, a wide range of masses could be generated.  
They may produce the heavy elements observed in the Lyman-alpha forest clouds 
at redshift z=2 to 4. They could  as well cause enough ionizing radiation
to reheat the universe at redshifts z$\ge$5, as it is required by the observations.  

However, in order to understand the role played by population III in the galactic evolution,
we have to know its initial
mass function (IMF).
Recently Nakamura and Umemura (1999)  
have obtained that the typical mass of Pop III is around 3M$_\odot$, that may grow 
by accretion up to 16M$_\odot$. Previously Yoshii and Saio (1986) found that the peak of the 
Pop III IMF is around 4-10M$_\odot$. Then it seams that intermediate mass stars were 
abundant in the early universe.

\begin{figure}
\epsfysize=6cm 
\epsfxsize=10cm 
\hspace{1.cm}\epsfbox{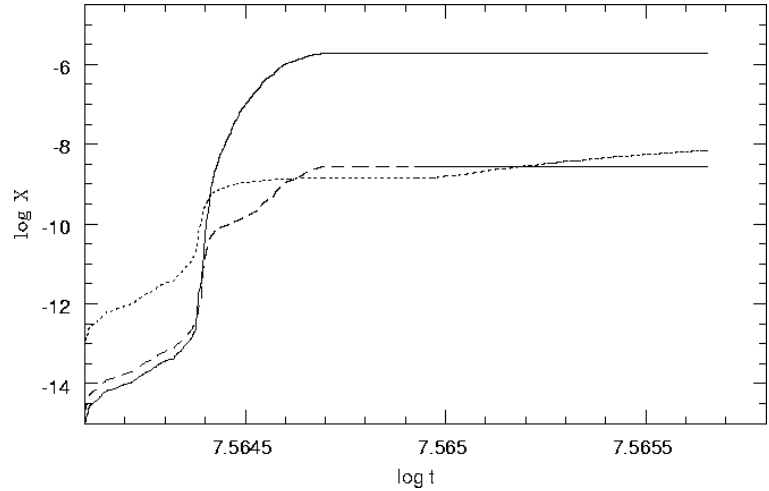} 
\caption[h]{Evolution of the surface mass fraction of $^{12}$C (solid line),
$^{14}$N (dashed line) and $^{16}$O (dotted line), in the 7 M$_\odot$.}
\end{figure}

\begin{figure}
\epsfysize=6cm 
\epsfxsize=10cm 
\hspace{1.cm}\epsfbox{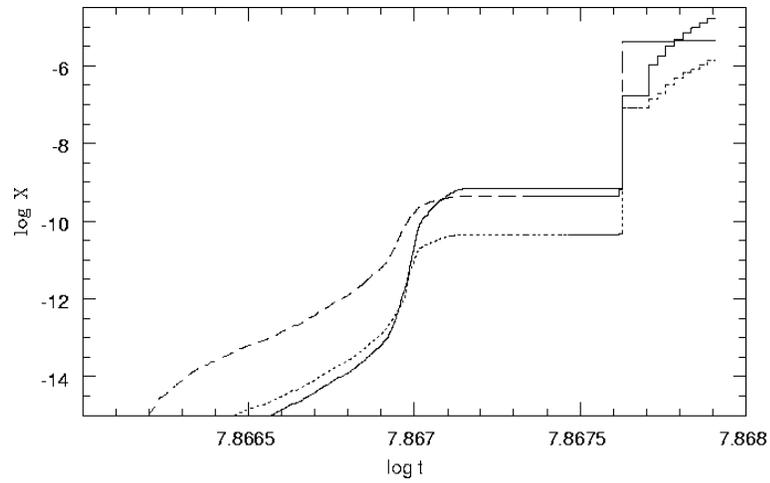} 
\caption[h]{Evolution of the surface mass fraction of $^{12}$C (solid line),
$^{14}$N (dashed line) and $^{16}$O (dotted line), in the 5 M$_\odot$.
The first episode of dredge-up
occurs during the E-AGB. The second one (around log t=7.8676) marks the onset of the TP-AGB
phase (see text).}
\end{figure}

Nevertheless just a single massive star exploding as 
a core collapse supernovae would increase the metallicity of a surrounded 
10$^{5}M_\odot$ 
cloud up to $Z\sim10^{-5}$. For this reason, most of the evolutionary calculations of
zero-metal stars concentrate on high masses. Some studies are also devoted to low mass
stars, since they would be still evolving at present time and, then, they could be observed.
In contrast, despite their potential importance in nucleosynthesis and chemical evolutions,
just few works have addressed the computation of the evolutionary properties
of Pop III intermediate mass stars (i.e. $3{\le}M/M_{\odot}{\le}8$).  

The only previous work in which the evolution of a zero metal intermediate mass star
is followed 
further than the early AGB (E-AGB)
is that of Chieffi and Tornamb\`e (1984). In particular they
studied the evolution of a Z=0,  5M$_\odot$ star.
Instead of the usual thermal pulses, they found 
that, after some small instabilities, both shells advance  
contemporaneously and experience a steady burning. The main reason is that due to the lack 
of CNO nuclei, the $3\alpha$ reaction must be active in the H shell to produce 
the necessary  amount of C for the CNO burning. If He burning occurs in the H external 
shell it will also occur contemporaneously in the inner and hotter He burning shell. 
Fujimoto et al. (1984) developed a semianalytical method to study the properties of the 
H and He burning shells. They found that if the H exhausted core mass is larger than
a critical value (which depends on the amount of CNO) a steady double shells burning occurs.
For Z=0 the critical core mass is 0.73 $M_{\odot}$. The larger the initial CNO abundance
the larger the critical core mass.

\begin{figure}
\epsfysize=6cm 
\epsfxsize=8cm 
\hspace{2cm}\epsfbox{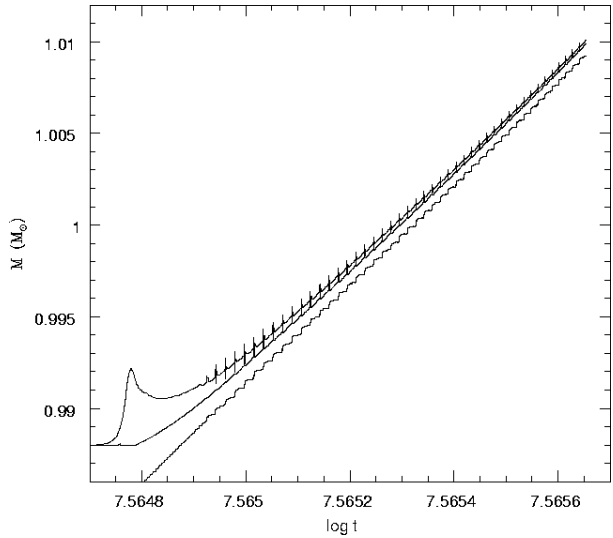} 
\caption[h]{Enlarged version of figure 1. The details of the TP-AGB phase are shown.}
\end{figure}
 
In this paper we firstly present the evolution of intermediate mass zero metal stars 
with masses over 5M$_\odot$. Contrary to the general picture of Fujimoto el al. (1984) 
these stars experience the TP-AGB phase, behaving as {\it normal} stars. 
The reason is that the Carbon abundance in the envelope at the end 
of the E-AGB phase is not the original 
one (i.e. Z=0). During the E-AGB the convective envelope penetrates and dredges up the $^{12}$C,   
previously produced by the $3\alpha$ reaction which operates within the H burning shell.  
This $^{12}$C provides the necessary  catalyst for an efficient CNO burning
and aloows the occurrence of the thermal pulses.
The 5M$_\odot$ is a limiting case.  The $^{12}$C enhancement of the envelope 
is not enough and the $3\alpha$ reactions must provide part of the total energy  
during most of the AGB lifetime.  
However, as it will be 
described later, after {\it some time}, two convective shell episodes are capable to newly dredge
up Carbon from the He core and finally the star enters the 
{\it normal} TP-AGB evolution.

\section{The models}
 
All the evolutionary models presented in this work have been computed by 
means of the latest version of the FRANEC
(Frascati RAphson Newton Evolutionary Code; release 4.7).  
We recall that the nuclear burning and the physical 
evolution are coupled and that a time dependent mixing scheme is adopted. The 
nuclear network includes 41 isotopes (269 reactions) for the H burning and 
26 isotopes (147 reactions) for the He burning. In addition a reduced set of nuclear species
and related reaction
have been added for the Carbon burning, namely 9 isotopes (8 reactions), just to
identify the value of $M_{up}$.
For a detailed description of the code see Chieffi, Limongi \& Straniero (1998) 
       
We present here the evolution of a 5, 6 and 7 M$_\odot$ (Z=0 and Y=0.23)
from the pre-main 
sequence up to the AGB. We have also followed the evolution of an 8 $M_\odot$
up to an off center Carbon ignition.
In this first investigation no mass loss has been assumed. 

\section {Pre-AGB evolution}
The qualitative behaviors of our models before the AGB phase does not substantially differ
from the ones already known (Ezer 1961, 1972, 1981; Ezer and Cameron 1971;
Wagner 1974; D'Antona and Mazzitelli 1982; Castellani, Chieffi and Tornamb\`e 1983; 
Tornamb\`e and Chieffi 1986; Cassisi and Castellani 1993; Cassisi, Castellani
and Tornamb\`e 1996). Let us just recall the main characteristic of the various phases.

During the pre-main sequence these stars contract
till the conditions for H burning via the p-p chain are achieved. The resulting pre-main
sequence lifetime is longer for the zero metal models, as compared 
with the same masses having larger metallicities (i.e. in which the CNO cycle can occur).
The ZAMS is located at higher L and T$_{eff}$. 
A small convective core develops (half the size of the one corresponding to a
{\it normal} metallicity star), but the H burning extends far outside this core
(it covers almost the 80\% of the total mass of the star). 
The H ignition does not stop the contraction and the temperature continues to rise, until
the $3\alpha$ reactions start and some Carbon is produced.
Note that only a small quantity of $^{12}C$ ($X_{C}\sim10^{-11}$) is sufficient to switch the 
H burning to the more energetic CNO cycle. 
As a consequence of this new (and more efficient) burning regime,
the local luminosity increases and
the convective core grows in mass. Near the end of the central H burning the convective core
disappears and the overall contraction occurs.
As usual, after the H exhaustion the central region of the star contracts again, 
until the central He
ignites and a new convective core develops. More outside the H burns in a thick shell.
During the major part of the He burning the star remains in the blue side of the HR diagram,
so that no dredge-up occurs. For this reason, these stars enter the AGB
with the original surface composition. 

\vspace{1cm} 
\begin{table}[htb]
\caption{}
\hspace{1.cm} 
\begin{tabular}{|l|c|c|c|c|c|}
\hline
\hline
M(M$_\odot$) & M$_{H}(M_\odot)$ & $^{4}$He & $^{12}$C & $^{14}$N & $^{16}$O \\
\hline
5.0 & 0.886 & 0.365 & $9.47\cdot10^{-10}$ & $4.29\cdot10^{-10}$ & $4.99\cdot10^{-11}$  \\
6.0 & 0.933 & 0.367 & $8.44\cdot10^{-8}$  & $8.29\cdot10^{-10}$ & $1.17\cdot10^{-10}$   \\
7.0 & 0.988 & 0.367 & $1.88\cdot10^{-6}$  & $1.41\cdot10^{-9}$  & $2.71\cdot10^{-9}$    \\
\hline
\hline
\end{tabular}
\end{table}

\begin{figure}
\epsfysize=6cm 
\epsfxsize=10cm 
\hspace{1.5cm}\epsfbox{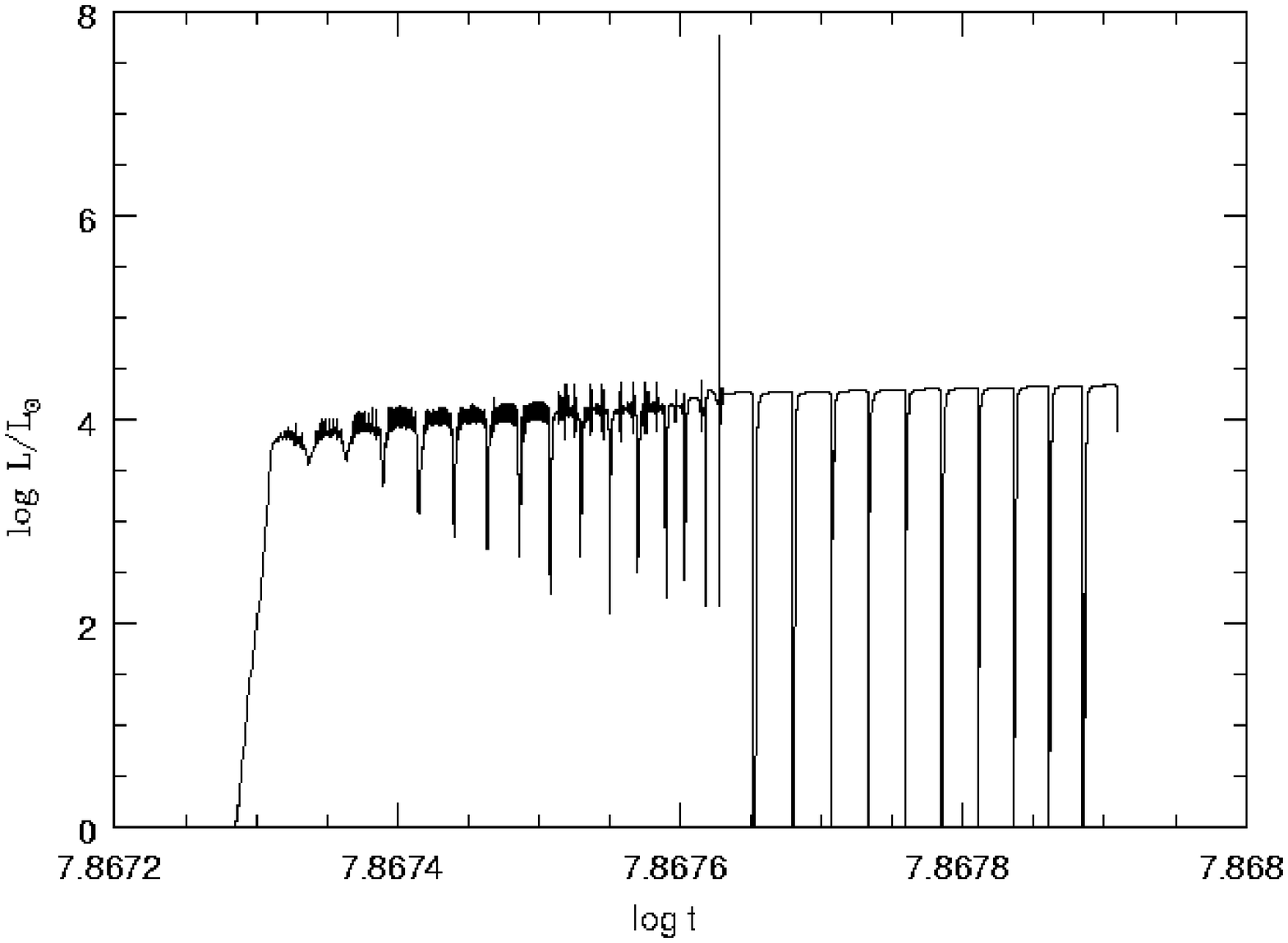} 
\caption[h]{Variations of the  H burning luminosity during the TP-AGB
of the 5 M$_\odot$.}
\end{figure}

\begin{figure}
\epsfysize=6cm 
\epsfxsize=10cm 
\hspace{1.5cm}\epsfbox{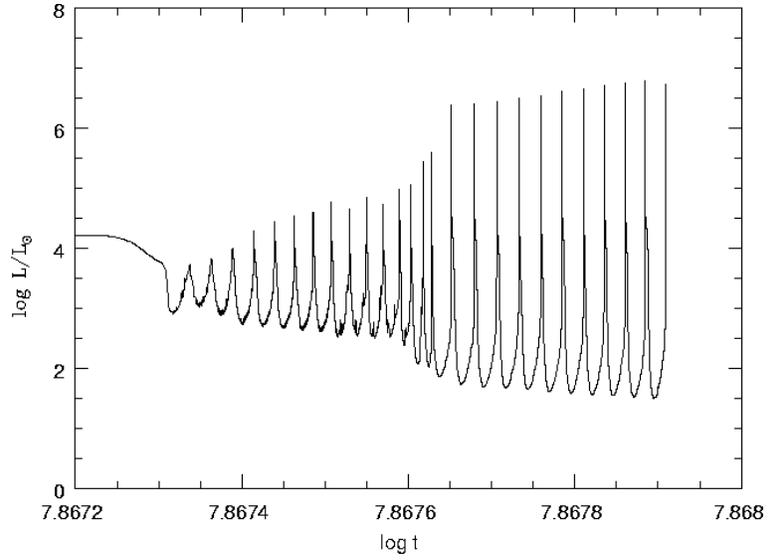} 
\caption[h]{Variations of the  He burning luminosity during the TP-AGB
of the 5 M$_\odot$.}
\end{figure}

\section{The Early AGB phase}

At the end of the central He burning these stars move in the red part of the HR diagram
and a convective envelope appears.
In figure 1  we show the variation of the locations of the H and He burning shells
as well as the location of the inner edge of the convective envelope, 
for the 7 $M_{\odot}$ models.  
Despite the similarity with the
classical second dredge-up, occurring in more metal rich stars,
in this case the modification
of the envelope composition presents important and peculiar characteristics.
It will play an important role in determine the further evolution. 
In table 1 we have summarized some properties of our models at the
end of the early AGB (E-AGB), namely (from column 1 to 6):
the total mass, the mass of the He core, 
the surface abundances (mass fraction) of $^{4}$He,  
$^{12}$C, $^{14}$N and $^{16}$O.

Note the very large amount of He brought to the surface (the initial one was 0.23).
This He was mainly produced during the extended central H burning
(see previous section).  At variance with {\it normal} stars for which
the total amount of CNO is not modified by the $2^{nd}$ dredge-up, 
here there is an important rise of all these elements.
In fact,  the surface Carbon abundance comes from the $3\alpha$,
which were active during the H-burning. Such a Carbon was subsequently partially
burned by the CNO, allowing a certain production of Oxygen and Nitrogen. 
In figure 2 and 3 we report the variation of the surface abundance of the CNO for
the 7 and 5 M$_\odot$, respectively.  

Before to describe the subsequent evolution let us remind that the 8 M$_\odot$
of Z=0 ignites Carbon. This is actually an off-center ignition, as usual
in degenerate core where, due to the significant thermal neutrino emission,
the maximum temperature does not coincides
with the center. We have follow part of this C
burning through the formation of an extended convective shell.
Thus we can conclude that for Pop III stars: $7\le M_{up}/M_{\odot}\le 8$.

\section{The advanced AGB evolution}

It is commonly believed that a zero metal star of intermediate mass 
does not experience the usual thermal
instabilities (thermal pulse or TP), which characterize the AGB evolution,
unless its core mass is lower than a critical value (Fujimoto et al. 1984).
This is certainly true if these stars can maintain the original (no metals) composition
in the envelope, so that the shell CNO burning cannot take place.
In our models of $M\ge6$ M$_\odot$, the surface abundance of CNO, after the $2^{nd}$ dredge up
(actually the first), is large enough to allow a {\it normal} CNO burning
and, in turn, to enter the TP-AGB phase. The case of the 7 $M_\odot$  is illustrated
in figure 4.

The evolutionary history of these thermally pulsing AGB stars
is similar to the one found
in more metal rich AGB stars. The quiescent H burning is broken by a series of recursive
and strong flash He burnings, which induce the formation of  extended convective shells.
After few pulses the envelope penetrates the region enriched with the products of the
CNO and $3\alpha$ reactions, so dredging up more CNO elements (see figure 2 and 3).
During the interpulse, the base of the convective envelope is generally too cool
for the hot bottom burning.
The most important difference with respect  to the {\it normal} AGB stars is the lack of 
iron seeds which prevents any s-elements production.  

On the contrary, in the 5M$_\odot$, the onset of the TP phase presents some peculiarities.
In such a case the amount of CNO nuclei left by the $2^{nd}$ dredge-up
is not sufficient to ensure an efficient H burning.
In figure 5 and 6 we report the variation of the H and He burning luminosities.
A first period is characterized by weak pulses. At the beginning of this period,
the $3\alpha$
luminosity peak is about $10^4 L_{\odot}$, then comparable with the H burning luminosity.
No convective shell forms. During the interpulses the He burning shell still
provides a non negligible fraction of the total nuclear energy.
However the strength of these weak pulses increases and, after about 9 of them,
a convective shell appears above the He burning region. Immediately after, another
convective episode, initially confined at the base of the H rich envelope, penetrates
the H/He discontinuity and overlaps the previous one (see figure 7).
\begin{figure}
\epsfysize=6cm 
\epsfxsize=8cm 
\hspace{2.cm}\epsfbox{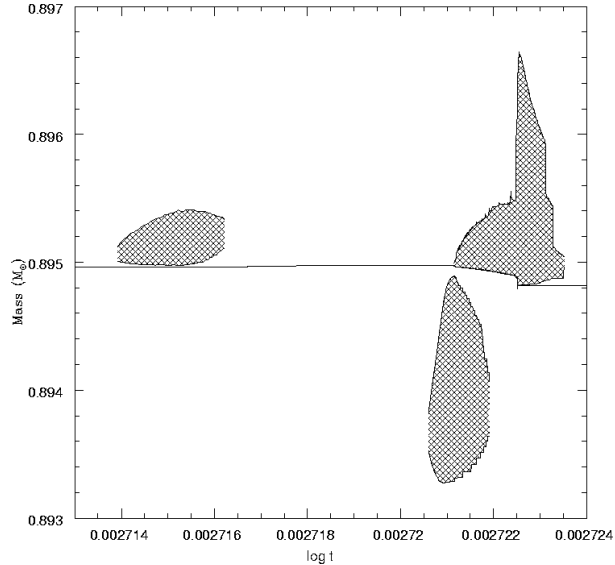} 
\caption[h]{Convective episodes after the $9^{th}$ weak TP of the 5 $M_{\odot}$,  the line
corresponds to the H/He discontinuity.}
\end{figure}

Then a lot of Carbon is dredged up and the H-burning experience a rather strong flash.
Later on the evolution proceeds as in the case of the more massive sequences.

\section{Conclusions}
 
By means of numerical models of the first generation of stars (Z=0, Y=0.23)
with masses greater than 5 M$_\odot$, we found that, 
during the AGB, the  
metal content of the envelope increases (up to $Z\sim10^{-5}$).
It is essentially enriched of primary Carbon, Nitrogen and Oxygen,
which come from the inner regions simultaneously processed by both the $3\alpha$
and the CNO cycle.
The surface Helium increases too, namely up to Y$\approx$0.37,
almost the double of the initial value.
We are now analysing in more details the nucleosynthesis occurring during the TP-phase
to derive a better estimation of the yields of Pop III stars.  
The result of this investigation will be presented in a forthcoming paper.
    
\acknowledgements

This work was partially supported by the MURST italian grant Cofin98,
by the MEC spanish grant PB96-1428, by the Andalusian grant FQM-108 and it
 is part of the ITALY-SPAIN
integrated action (MURST-MEC agreement) HI98-0095 .

\end{document}